\documentstyle[aps,12pt]{revtex}

\begin{document}
\title{Realization of Probabilistic Identification and Clone of Quantum-States II
Multiparticles System }
\author{Chuan-Wei Zhang, Chuan-Feng Li, Guang-Can Guo\thanks{%
Email: gcguo@ustc.edu.cn}}
\address{Department of Physics and Laboratory of Quantum Communication and Quantum\\
Computation,\\
University of Science and Technology of China,\\
Hefei 230026, P. R. China\vspace*{0.3in}}
\maketitle

\begin{abstract}
\baselineskip24ptWe realize the probabilistic cloning and identifying linear
independent quantum states of multi-particles system, given prior
probability, with universal quantum logic gates using the method of unitary
representation. Our result is universal for separate state and entanglement.
We also provide the realization in the condition given $M$ initial copies
for each state.

 \newpage\ 
\end{abstract}

\baselineskip24pt

\section{BACKGROUND}

The development of quantum information theory [1] has drawn attention to
fundamental questions about what is physically possible and what is not. An
example is the quantum no-cloning theorem [2,3], which asserts that unknown
pure states can not be reproduced or copied by any physical means. Wootters
and Zurek [2] shows that the cloning machines violates the quantum
superposition principles, which applies to a minimum total number of three
states, and hence does not rule out the possibility of cloning two
non-orthogonal states. Yuen and D'Ariano [4,5] show that a violation of
unitarity makes the cloning of two non-orthogonal states impossible.
Recently, Barnum et al. [6] have extended such results to the case of mixed
states, and it was shown that two non-commuting mixed states can not be
broadcast onto two separate quantum systems, even when the states need only
be reproduced marginally. More recently, Koashi and Imoto [7] extended the
standard no-cloning theorem to the case of a subsystem correlated to others
and derive a necessary and sufficient condition for two pure states, each
entangled in two remote systems, to be clonable by the sequential access to
the two systems. Identification and clone have close relation and also we
can not identify an arbitrary unknown state. Since quantum states can not be
cloned and identified faithfully, recently, the inaccurate coping and
discriminating of quantum states have aroused great interest.

The inaccurate coping and discriminating of quantum states can be divided
into two main categories: universal and state-dependent. The first category
is aimed at the approximate clone or identification of an arbitrary state
[8-13], of which the figuration of merit is the local fidelity. The second
category is designed to identify and clone only a finite number of states
and here we can identify two sub-categories. Approximate state-dependent
cloning machines deterministically generate approximate clones of state
belonging to a finite set [14-16]. Exact state-dependent cloning machines
clone a set of non-orthogonal but liner-independent pure states with the
limited success probability [17-20]. This job is done by Duan and Guo and
they have found the maximum success probability if these copies are required
to be exact clones. Duan and Guo have provided a new possibility to copy
states and demonstrated that these clones can get to the maximum success
probability which is non-zero. More recently, we [21] constructed general
discriminating strategies of the quantum states secretly chosen from a
certain set , given initial $M$ copies of each state, and obtained the
matrix inequality, which describe the bound between the maximum probability
of correctly determining and that of error . We [22] also obtained the
unitary representations and Hamiltonians for the most general quantum
probabilistic clone and identification of $n$ linear independent states with
the given probability. For application we realized it in $2$-state system
with universal quantum logic gates under the condition that the maximum
probability is chosen.

In this paper we will apply our method, which was provided in [22], and
realize the probabilistic cloning and identifying linear independent quantum
states of multi-particles system with universal logic gates. Entanglement is
a special state of multi-particles system, so we realize the probabilistic
clone and identification of entanglement states in actual. In Section II, we
will introduce our method in [22] and some results about universal logic
gates. In Section III, we execute this method on $2$-particles system and
realize the probabilistic cloning and identifying four linear independent
states of two particles. It is obvious that entangled states of two
particles are only special cases of our result. We generalize our result to $%
n-$particles system and we also provide the realization of the probabilistic
clone and identification if we give $M$ initial copies for each state.

\section{INTRODUCTION}

We begin our work by introducing the result in [22], which is also the basis
method used in this article. Consider the question of probabilistic
identification if we have $M$ initial copies of each state. We are concerned
with a set of $n$ linear-independent quantum states$\left\{ \left| \psi
_i\right\rangle ,\text{ }i=1,2,~\cdots ,n\right\} $, which span an $n$%
-dimensional Hilbert space ${\cal H}$. We use a general unitary-reduction
operation and get

\begin{equation}
\hat U\left| \psi _i\right\rangle ^{\otimes M}\left| P_0\right\rangle =\sqrt{%
\gamma _i}\left| \widetilde{\varphi _i}\right\rangle \left| P_1\right\rangle
+\sum_jc_{ij}\left| \alpha _j\right\rangle \left| P_0\right\rangle \text{,} 
\eqnum{2.1}
\end{equation}
where $\gamma _i$ is the probability of success, $\left| \psi
_i\right\rangle ^{\otimes M}=\left| \psi _i\right\rangle _1\left| \psi
_i\right\rangle _2\cdots \left| \psi _i\right\rangle _M$, $\left| \psi
_i\right\rangle _k$ is the $k$-th cope of state$\left| \psi _i\right\rangle $%
. For $\left| \psi _i\right\rangle ^{\otimes M}$ are liner-independent, they
span a $n$-dimensional Hilbert space. We denote $\left\{ \left| \alpha
_j\right\rangle ,\text{ }j=1,2,\cdots ,n\right\} $ as the orthogonal basis
states of that space. $\left\{ \left| \widetilde{\varphi _i}\right\rangle
,i=1,2,\cdots ,n\right\} $ is a set of orthogonal states in $n^M$%
-dimensional Hilbert space, whose orthogonal basis states are $\{\left|
\varphi _{i_1}\right\rangle _1\left| \varphi _{i_2}\right\rangle _2\cdots
\left| \varphi _{i_M}\right\rangle _M$, $i_j=1,2,\cdots ,n$, $j=1,2,\cdots
,M\}$, where $\left\{ \left| \varphi _i\right\rangle ,\text{ }i=1,2,\cdots
,n\right\} $ are the orthogonal basis states of ${\cal H}$. $\left\{ \left|
P_0\right\rangle ,\text{ }\left| P_1\right\rangle \right\} $ are orthogonal
basis states of the probe system.

The unitary evolution $\hat U$ exists if and only if

\begin{equation}
X^{(M)}=\Gamma +CC^{+}\text{,}  \eqnum{2.2}
\end{equation}
where $X^{(M)}=\left[ \left\langle \psi _i\mid \psi _j\right\rangle
^M\right] $, $\Gamma =diag\left( \gamma _1,\gamma _2,...,\gamma _n\right) $, 
$C=\left[ c_{ij}\right] _{n\times n}$.

Equation (2.2) is equivalent to the following inequality 
\begin{equation}
X^{(M)}-\Gamma \geq 0.  \eqnum{2.3}
\end{equation}

Inequality (2.3) determines maximum probability of success.

We choose the probability matrix $\Gamma $ and give a set of parameters that
are determined by 
\begin{equation}
I-C^{+}X^{(M)-1}C=Vdiag(m_1,m_2,\cdots ,m_n)V^{+}.  \eqnum{2.4}
\end{equation}

In the condition of Eq. (2.4) and the orthogonal basis states $\left\{
\left\{ \left| \alpha _i\right\rangle \left| P_0\right\rangle \right\} ,%
\text{ }\left\{ \left| \widetilde{\varphi _i}\right\rangle \left|
P_1\right\rangle \right\} ,\text{ }i=1,2,\cdots n\right\} $, we can
represent $U$ as

\begin{equation}
U=\widetilde{V}\left( 
\begin{array}{cc}
F & -E \\ 
E & F
\end{array}
\right) \tilde V^{+}\text{,}  \eqnum{2.5}
\end{equation}
where $E=diag(\sqrt{m_1},\sqrt{m_2},...,\sqrt{m_n})$, $\left. F=diag(\sqrt{%
1-m_1},\sqrt{1-m_2},...\right. $, $\sqrt{1-m_n})$, $\widetilde{V}=diag(V,V)$.

Now we shall be concerned with clone in the circumvent that if we have $M$
initial copies and wish to obtain $N$ copies, where $N>M$. We use the
unitary evolution as following 
\begin{equation}
U\left| \psi _i\right\rangle ^{\otimes M}\left| \varphi _1\right\rangle
^{\otimes (N-M)}\left| P_0\right\rangle =\sqrt{\gamma _i}\left| \psi
_i\right\rangle ^{\otimes N}\left| P_i\right\rangle +\sum_jc_{ij}\left|
\alpha _j\right\rangle \left| \varphi _1\right\rangle ^{\otimes (N-M)}\left|
P_0\right\rangle \text{,}  \eqnum{2.6}
\end{equation}
where $\left| P_0\right\rangle $ and $\left| P_i\right\rangle $ are
normalized states of the probe system(not generally orthogonal) but $\left|
P_0\right\rangle $ is orthogonal with each of $\left| P_i\right\rangle $.
The using of $\left| \psi _i\right\rangle ^{\otimes N}$ has the similar
definition with $\left| \psi _i\right\rangle ^{\otimes M}$ and the basis
states of space $\left\{ \left| \psi _i\right\rangle ^{\otimes N}\left|
P_i\right\rangle ,\text{ }i=1,...,n\right\} $ are $\left\{ \left| \beta
_j\right\rangle ,\text{ }j=1,2,\cdots ,n\right\} .$

$U$ exists if and only if

\begin{equation}
X^{(M)}=\sqrt{\Gamma }X_P^{(N)}\sqrt{\Gamma }+CC^{+}\text{,}  \eqnum{2.7}
\end{equation}
where $X_P^{(N)}=\left[ \left\langle \psi _i|\psi _j\right\rangle
^N\left\langle P_i|P_j\right\rangle \right] .$

Eq. (2.7) is equivalent to the following 
\begin{equation}
X^{(M)}-\sqrt{\Gamma }X_P^{(N)}\sqrt{\Gamma }\geq 0.  \eqnum{2.8}
\end{equation}

Given probability matrix $\Gamma $, under the condition (2.4) but the
different orthogonal basis vectors $\left\{ \left\{ \left| \alpha
_i\right\rangle \left| \varphi _1\right\rangle ^{\otimes (N-M)}\left|
P_0\right\rangle \text{, }i=1,2,\cdots n,\right\} ,\;\left\{ \left| \beta
_i\right\rangle ,\;i=1,2,...,n\right\} \right\} $, we represent $U$ as the
same form as that in Eq.(2.5).

With the concrete form of the unitary representation of the evolution in
cloning machine, we can realize that probabilistic clone and identification
with universal quantum logic gates, for Deutsch [23] have proved that any
unitary evolution could be executed with universal quantum logic gates, i.e. 
$\Lambda _m(U)$. A. Barenco et al. [24] has demonstrated that any arbitrary $%
\Lambda _m(U)$ could be realized by the combination of a set of gates of all
one-bit quantum gates $[U(2)]$ and the two-bit Control-NOT gate(that maps
Boolean values $(x,y)$ to $(x,x\oplus y)$). So our work is to realize our
result with gate $\Lambda _m(U)$ .

In the following part of this section we will introduce some basis ideas and
notation about universal quantum logic gates. For any unitary matrix $%
U=\left( 
\begin{array}{ll}
u_{00} & u_{01} \\ 
u_{10} & u_{11}
\end{array}
\right) $ and $m\in \{0,1,2,...\}$, the matrix corresponding to the $(m+1)$%
-bit operator $\Lambda _m(U)$ is $diag(I_{2^m},U)$, where the basis states
are lexicographically ordered, i.e., $\left| 000\right\rangle ,\;\left|
001\right\rangle ,...,\;\left| 111\right\rangle $.

When $U=\left( 
\begin{array}{ll}
0 & 1 \\ 
1 & 0
\end{array}
\right) $, $\Lambda _m(U)$ is so called Toffoli gate [23] with $m+1$ input
bits. For a general $U$, $\Lambda _m(U)$ can be regarded as a generalization
of the Toffoli gate, which, on the $m+1$ input bits, applies $U$ to the $%
(m+1)$--th bit if and only if the other $m$ bits all are on $1$.

\section{REALIZATION FOR MULTI-PARTICLES SYSTEM}

We begin our discussion with introducing two lemma which tell us how to
decompose a general unitary matrix to the products of the matrixes of the
form $\Lambda _m(U)$.

{\sl LEMMA 1:} Any unitary matrix $U=\left[ u_{ij}\right] _{n\times n}$ can
be decomposed into the following form 
\begin{equation}
U=\left( \prod_{t=1}^{n-1}\prod_{l=t+1}^nA_{tl}\right) \left(
\prod_{k=1}^nB_k\right) ,  \eqnum{3.1}
\end{equation}
where $A_{tl}=[a_{ij\text{ }}^{(tl)}]_{n\times n}$ $=P_{t,n-1}P_{l,n}\hat A(%
\hat u_{tl})P_{l,n}^{+}P_{t,n-1}^{+}$, $P_{ij}$ is the unitary matrix which
interchange between the $i$-th row and $j$-th row of $\hat A(\hat u_{tl})$
and $P_{ij}^{+}$ interchange the $i$-th collum and $j$-th collum of $\hat A(%
\hat u_{tl})$. It is obvious that $P_{ij}=P_{ij}^{+}$. $\hat A(\hat u%
_{tl})=diag(1,1,...,1,\hat u_{tl})$, where $\hat u_{tl}$ is a $2\times 2$
unitary matrix. $B_k=P_{k,n}\hat B(\exp (i\alpha _k))P_{k,n}$, where $\hat B%
(\exp (i\alpha _k))=diag(1,1,...,1,\exp (i\alpha _k))$.

The meaning of this decomposition in math is that we use some unitary matrix
to transfer the non-diagonal elements of $U$ into $0$. the decomposition is
stylized and can be easily finished by any classical computer. $A_{tl}$ and $%
B_k$ are completely determined by $U$. When $n=2^{m+1}$, we find $\hat A(%
\hat u_{tl})=\Lambda _m(\hat u_{tl})$, $\hat B(\exp (i\alpha _k))=\Lambda
_m(diag(1,\;\exp (i\alpha _k))$. We suppose in the subspace which is spanned
by basis states $\left| x_1^i,\;x_2^i,...,\;x_{m+1}^i\right\rangle $ and $%
\left| x_1^j,\;x_2^j,...,\;x_{m+1}^j\right\rangle $, for all $%
x_k^i,\;x_k^j\in \left\{ 0,1\right\} ,\;k=1,2,...,m+1$. When $i\neq j$,
there must exist $k$ making $x_k^i\neq x_k^j$. Denote the minimum value of $%
k $ still as $k$, we generally assume $x_k^i=1$, then $x_k^j=0$. For $%
k<s\leq n $, if $x_s^i\neq x_s^j$, we execute the $k$--th bit control $%
\sigma _x^s$ operation on the $s$--th bit (C-NOT gate). Then for $1\leq
h\leq n$, when $h\neq k$ and $x_h^i=x_h^j=0$, we execute $\sigma _x^h$ on
the $s$--th bit. At last we interchange the input sequence of the $k$--th
bit and the $(m+1)$--th bit. With all the above operations, we transfer the
operation $P_{ij}$ into $\Lambda _m(\sigma _x)$.

With above lemma, we can express a unitary evolution as the products of some
controlled unitary operations.

In this section we first limit our discussion in two particles, which is
four states system, and suppose $M=1$, $N=2$. For general $n$ particles and $%
M$, $N$, we will discuss that in the last part. The realization of
probabilistic clone and identification for two particles system has
important value in practice, for the entanglement of two particles is a
special state of two particles system. We assume each particle reduce to
either of two states $\left| 0\right\rangle $ or $\left| 1\right\rangle $
after we execute measurement $\sigma _z$ on them. So the states of the two
particles span a $4$-dimension Hilbert space ${\cal H}$, whose orthogonal
basis states are $\left\{ \left| \phi _1\right\rangle _{1,2}=\left|
00\right\rangle _{1,2},\;\left| \phi _2\right\rangle _{1,2}=\left|
01\right\rangle _{1,2},\;\left| \phi _3\right\rangle _{1,2}=\left|
10\right\rangle _{1,2},\;\left| \phi _4\right\rangle _{1,2}=\left|
11\right\rangle _{1,2}\right\} $. Consider four linear independent quantum
states $\left\{ \left| \psi _i\right\rangle _{1,2},\;i=1,2,3,4\right\} $. We
express each state on the above basis states as 
\begin{equation}
\left| \psi _i\right\rangle _{1,2}=\sum_jt_{ij}\left| \phi _j\right\rangle
_{1,2},  \eqnum{3.3}
\end{equation}
where $t_{ij}$ satisfy $\sum_j\left| t_{ij}\right| ^2=1$. Define $%
T=[t_{ij}]_{4\times 4}$. The determinant of $T$ should be non-zero because $%
\left| \psi _i\right\rangle _{1,2}$ are linear independent. Contrast with
the situation of $3$-dimension, we give following lemma

{\sl LEMMA 2: }For any four states $\left\{ \left| \psi _i\right\rangle
_{1,2},\;i=1,2,3,4\right\} $ in Hilbert space ${\cal H}$, there exist a
unitary operator $U_0$ to make 
\begin{equation}
U_0\left( \left| \psi _1\right\rangle _{1,2},\;\left| \psi _2\right\rangle
_{1,2},\;\left| \psi _3\right\rangle _{1,2},\;\left| \psi _4\right\rangle
_{1,2}\right) =\left( \left| \phi _1\right\rangle _{1,2},\;\left| \phi
_2\right\rangle _{1,2},\;\left| \phi _3\right\rangle _{1,2},\;\left| \phi
_4\right\rangle _{1,2}\right) \widetilde{T},  \eqnum{3.4}
\end{equation}
where $\widetilde{T}=\left( 
\begin{array}{cccc}
1 & e^{i\mu _2^1}\cos \theta _2^{(1)} & e^{i\mu _3^1}\cos \theta
_3^{(1)}\cos \theta _3^{(2)} & e^{i\mu _4^1}\cos \theta _4^{(1)}\cos \theta
_4^{(2)}\cos \theta _4^{(3)} \\ 
0 & \sin \theta _2^{(1)} & e^{i\mu _3^2}\cos \theta _3^{(1)}\sin \theta
_3^{(2)} & e^{i\mu _4^2}\cos \theta _4^{(1)}\cos \theta _4^{(2)}\sin \theta
_4^{(3)} \\ 
0 & 0 & \sin \theta _3^{(1)} & e^{i\mu _4^3}\cos \theta _4^{(1)}\sin \theta
_4^{(2)} \\ 
0 & 0 & 0 & \sin \theta _4^{(1)}
\end{array}
\right) $ and $0\leq \theta _i^{(1)}<\pi $, $0\leq \theta _i^{(j)}<2\pi $, $%
0\leq \mu _i^j<2\pi $, if $\left\{ \left| \psi _i\right\rangle
_{1,2},\;i=1,2,3,4\right\} $ are linear independent, there must be $\theta
_i^{(1)}>0$. $\widetilde{T}$ is determined by 
\begin{equation}
T^{+}T=\tilde T^{+}\tilde T.  \eqnum{3.5}
\end{equation}

So in the following we only consider the probabilistic clone and
identification of states with the form as $\left| \tilde \psi
_i\right\rangle _{1,2}=U_0\left| \psi _i\right\rangle _{1,2}$. With {\sl %
LEMMA 1 }we can decompose $U_0$ to basis operation.

\subsection{Realization of Probabilistic Identification}

Given a set of linear independent quantum states of two particles system $%
\left\{ \left| \tilde \psi _i\right\rangle _{1,2},\;i=1,2,3,4\right\} $,
which have the forms we have introduced above. We use a general
unitary-reduction operation as that in Eq.(2.1) to execute the probabilistic
identification, where we let $M=1$. The maximum identification probability
is determined by Ineq.(2.3). Denote $\left| \alpha _i\right\rangle =\left|
\phi _i\right\rangle $ and $\left| \tilde \varphi _i\right\rangle =\left|
\phi _i\right\rangle $. On the given prior maximum probability matrix $%
\Gamma $ and orthogonal basis states $\left\{ \left\{ \left| \phi
_i\right\rangle _{1,2}\left| P_0\right\rangle ,\;i=1,2,3,4\right\} ,\text{ }%
\left\{ \left| \phi _j\right\rangle _{1,2}\left| P_1\right\rangle
,\;j=1,2,3,4\right\} \right\} $, $U$ can be determined by Eq.(2.5), where $%
\widetilde{V}$, $F$, $E$ are determined by Eq.(2.4). $\widetilde{V}$
represent the operate $\hat V_{1,2}\hat I_P$, where the matrix corresponding
to the operator $\hat V_{1,2}$ on the basis states $\left\{ \left| \phi
_i\right\rangle _{1,2}\right\} $ is $V$, $\hat I_P$ represent unity operator
of probe system. On new orthogonal basis states $\left\{ \left\{ \left| \phi
_i\right\rangle _{1,2}\left| P_0\right\rangle ,\;\left| \phi _i\right\rangle
_{1,2}\left| P_1\right\rangle \right\} ,\;i=1,2,3,4\right\} $, we express $%
S=\left( 
\begin{array}{cc}
F & -E \\ 
E & F
\end{array}
\right) $ as $S=diag(K_1,K_2,K_3,K_4)$, where $K_i=\left( 
\begin{array}{cc}
\sqrt{1-m_i} & -\sqrt{m_i} \\ 
\sqrt{m_i} & \sqrt{1-m_i}
\end{array}
\right) $. So we obtain 
\begin{equation}
\hat S=\sigma _x^1\sigma _x^2\Lambda _2(K_1)\sigma _x^2\sigma _x^1\cdot
\sigma _x^1\Lambda _2(K_2)\sigma _x^1\cdot \sigma _x^2\Lambda _2(K_3)\sigma
_x^2\cdot \Lambda _2(K_4).  \eqnum{3.6}
\end{equation}

We have showed in {\sl LEMMA 1 }that the unitary evolution $U_0$ and $\hat V%
_{1,2}$ can be decomposed into the product of basis operations such as C-NOT
and $\Lambda _2(\hat u)$. The decomposition of $\Lambda _2(\hat u)$ has been
completed by Barenco et al. [24]. For C-NOT gate we adopt the schematic
convention of Barenco et al. [24] by representing the transformations of the
target and control qubits by $\oplus $ and $\bullet $ respectively. So we
print the realization of quantum gates here

\[
\text{Figure 1.The networks of probabilistic identification.} 
\]
where the S-gate is illustrated in Fig. 2 
\[
\text{Figure 2.The S-gate realization.} 
\]

\subsection{Realization of Probabilistic Clone}

We still use a general unitary-reduction operation as that in Eq. (2.6) to
execute the probabilistic clone, where we let $M=1$, $N=2$. The maximum
clone probability is determined by Ineq.(2.8), where we let $\left\{ \left|
P_i\right\rangle =\left| P_1\right\rangle ,\;i=2,3,4\right\} $. We denote $A$
represent the system of the two particles $1,2$ and $B$ for $1^{^{\prime
}},2^{^{\prime }}$. To obtain the basis states $\left| \beta _j\right\rangle 
$, we first introduce a new operation which transfer all of the information
describing state $\left| \tilde \psi _i\right\rangle ^{\otimes N}$ the $N$
clone are in into one state of two particles. This operation must itself be
performed using only pairwise and local interactions. Denote vector $\theta
=\left\{ \theta _i^{(j)},\;\mu _i^j,\;2\leq i\leq 4,\;1\leq j\leq
i-1\right\} $, $\xi $, $\eta $ are denoted as same. This operation acts as
follows: 
\begin{equation}
D(\theta ,\xi )\left| \tilde \psi _i(\theta )\right\rangle _A\left| \tilde 
\psi _i(\xi )\right\rangle _B=\left| \tilde \psi _i(\eta )\right\rangle
_A\left| 00\right\rangle _B.  \eqnum{3.7}
\end{equation}

The operator $D(\theta ,\xi )$ is unitary, so we must have 
\begin{equation}
X(\theta ,\xi )=\tilde T^{+}(\eta )\tilde T(\eta ),  \eqnum{3.8}
\end{equation}
where $X(\theta ,\xi )=\left[ _A\left\langle \tilde \psi _i(\theta )\right|
_B\left\langle \tilde \psi _i(\xi )\right| \left| \tilde \psi _j(\theta
)\right\rangle _A\left| \tilde \psi _j(\xi )\right\rangle _B\right]
_{4\times 4}$. For the special form of $\left| \tilde \psi _i\right\rangle $%
, together with $0<\theta _i^{(1)}<\pi $, $0\leq \theta _i^{(j)}<2\pi $, $%
0\leq \mu _i^j<2\pi $, suffices to determine $\eta $ uniquely through
Eq.(3.8).

To obtain an explicit expression for the operator $D(\theta ,\xi )$, we must
first specify how it transforms states in the subspace orthogonal to that
spanned by $\left| \tilde \psi _i(\theta )\right\rangle _A\left| \tilde \psi
_i(\xi )\right\rangle _B$. A natural completion of the description of $%
D(\theta ,\xi )$ suggests itself if we take the sum and difference of the
four equation in Eq.(3.7), giving

\begin{equation}
D^{-1}(\theta ,\xi )\left\{ \left| \phi _i\right\rangle _A\left|
00\right\rangle _B,\;i=1,2,3,4\right\} =\left\{ \left| \phi _i\right\rangle
_A\left| \phi _j\right\rangle _B,\;i,j=1,2,3,4\right\} G\tilde T^{-1}(\eta )%
\text{,}  \eqnum{3.9}
\end{equation}
where $G_{16\times 4}$ is the matrix representation of states $\left\{
\left| \tilde \psi _i(\theta )\right\rangle _A\left| \tilde \psi _i(\xi
)\right\rangle _B\right. $, $\left. i=1,2,3,4\right\} $ on the basis states $%
\left\{ \left| \phi _i\right\rangle _A\left| \phi _j\right\rangle
_B,\;i,j=1,2,3,4\right\} $, which are lexicographically ordered, i.e., $%
\left| 0000\right\rangle ,\left| 0001\right\rangle ,...,\left|
1111\right\rangle $, of Hilbert space ${\cal H\otimes H}$. We denote $G%
\tilde T^{-1}(\eta )=\left( \omega _1,\omega _5,\omega _9,\omega
_{13}\right) $. It is obvious that the states $\left| \omega _i\right\rangle 
$ that $\omega _i$ represent are orthogonal and in the space spanned by $%
\left\{ \left| \tilde \psi _i(\theta )\right\rangle _A\left| \tilde \psi
_i(\xi )\right\rangle _B,\;i=1,2,3,4\right\} $. So we can choose states $%
\left\{ \left| \omega _j\right\rangle ,\;1\leq j\leq 16,\;j\notin \left\{
1,5,9,13\right\} \right\} $ in the subspace orthogonal to that spanned by $%
\left\{ \left| \omega _i\right\rangle ,\;i=1,5,9,13\right\} $ and denote $%
\tilde G^{-1}=\left( \omega _1,\omega _2,...,\omega _{16}\right) $. With
Eq.(3.9), we let 
\begin{equation}
D^{-1}(\theta ,\xi )\left\{ \left| \phi _i\right\rangle _A\left| \phi
_j\right\rangle _B\right\} =\left\{ \left| \phi _i\right\rangle _A\left|
\phi _j\right\rangle _B\right\} \tilde G^{-1}.  \eqnum{3.10}
\end{equation}
So we represent $D(\theta ,\xi )$ on the basis orthogonal states $\left\{
\left| \phi _i\right\rangle _A\left| \phi _j\right\rangle _B\right. $, $%
\left. i,j=1,2,3,4\right\} \;$as $\tilde G$. We find all the above steps for
obtaining the unitary representation of $D(\theta ,\xi )$ can be easily
performed with classical computer and the procedure is stylized.

Denote $\left. \left| \alpha _i\right\rangle =\left| \phi _i\right\rangle 
\text{,}\;\left| \varphi _1\right\rangle =\left| \phi _1\right\rangle
\right. $, $\left| \beta _i\right\rangle =D^{-1}(\theta ,\theta )\left| \phi
_i\right\rangle _A\left| \phi _1\right\rangle _B\left| P_1\right\rangle $.
So we give the orthogonal basis states as $\left\{ \left\{ \left| \phi
_i\right\rangle _A\left| \phi _1\right\rangle _B\left| P_0\right\rangle
\right\} ,\;\left\{ D^{-1}(\theta ,\theta )\left| \phi _i\right\rangle
_A\left| \phi _1\right\rangle _B\left| P_1\right\rangle \right\} \right. $, $%
\left. i=1,2,3,4\right\} $. We introduce two new gates called Control-$%
D(\theta ,\xi )$ and Control-$D^{+}(\theta ,\xi )$, whose function is that
we execute $D(\theta ,\xi )$ or $D^{+}(\theta ,\xi )$ operates on the two
target states when the control qubit on $\left| 1\right\rangle $, otherwise
we do not make influence on the two target states. Then we can transfer the
above orthogonal basis vectors into the new orthogonal basis states $\left\{
\left\{ \left| \phi _i\right\rangle _A\left| \phi _1\right\rangle _B\left|
P_0\right\rangle \right\} ,\;\left\{ \left| \phi _i\right\rangle _A\left|
\phi _1\right\rangle _B\left| P_1\right\rangle \right\} ,\;i=1,2,3,4\right\} 
$ with the C-$D(\theta ,\theta )$ gate and the inverse with the C-$%
D^{+}(\theta ,\theta )$ gate, where we take $P$ as the control qubit, and $%
A,B$ as the target states.

On the new orthogonal basis vectors, with the given maximum clone
probability matrix $\Gamma $, combine with Eq.(2.4) and Eq.(2.7), $U$ can be
obtained through Eq.(2.5), where $\widetilde{V}$, $E$, $F$ are determined by
Eq.(2.4) . $\widetilde{V}$ represent the operate $\hat V_A\left| \phi
_1\right\rangle _{BB}\left\langle \phi _1\right| \hat I_P$, where the matrix
corresponding to the operator $\hat V_A$ on the basis states $\left\{ \left|
\phi _i\right\rangle _A\right\} $ is $V$, $\hat I_P$ represent unity
operator of probe system, $\left| \phi _1\right\rangle _{BB}\left\langle
\phi _1\right| $ means only when $B$ is on $\left| \phi _1\right\rangle _B$
we can execute the operator $\hat V_A$ on system $A$. On new orthogonal
basis states $\left\{ \left\{ \left| \phi _i\right\rangle _A\left|
P_0\right\rangle ,\;\left| \phi _i\right\rangle _A\left| P_1\right\rangle
\right\} \otimes \left| \phi _1\right\rangle _B,\;i=1,2,3,4\right\} $, we
express $S=\left( 
\begin{array}{cc}
F & -E \\ 
E & F
\end{array}
\right) $ as $S=diag(K_1,K_2,K_3,K_4)$, where $K_i=\left( 
\begin{array}{cc}
\sqrt{1-m_i} & -\sqrt{m_i} \\ 
\sqrt{m_i} & \sqrt{1-m_i}
\end{array}
\right) $. So we obtain 
\begin{equation}
\hat S=\sigma _x^1\sigma _x^2\Lambda _2^{AP}(K_1)\sigma _x^2\sigma _x^1\cdot
\sigma _x^1\Lambda _2^{AP}(K_2)\sigma _x^1\cdot \sigma _x^2\Lambda
_2^{AP}(K_3)\sigma _x^2\cdot \Lambda _2^{AP}(K_4)\left| \phi _1\right\rangle
_{BB}\left\langle \phi _1\right| ,  \eqnum{3.11}
\end{equation}
where $\left| \phi _1\right\rangle _{BB}\left\langle \phi _1\right| $ means
only when $B$ is on $\left| \phi _1\right\rangle _B$ we can execute the
operator $\hat S$ on system $A$ and $P$. With all the above operator, we
realize the probabilistic clone of the states of two particles system. With
the concrete form of $D(\theta ,\xi )$, $\hat V_A$, $S$ and {\sl LEMMA 1},
we can decomposed these operators into the products of some basis operators
such as C-NOT and controlled unitary operators. We print the realization
with quantum logic gates as here

\[
\text{Figure 3. The networks of probabilistic clone.} 
\]

\subsection{General Results for Given $M$ Initial Copies}

In this part we will give the results in the situation if we have $M$
initial copies and want to identify them or clone $N$ copies. The methods
are similar. We first denote the operator which transfer the information of
the possible states of the particles system $A_{j+1}$ to those of the
particles system $A_j$ by $D_j(\theta ,\xi )$. We define $D_M=D_1\left(
\theta ,\xi _{M-1}\right) D_2\left( \theta ,\xi _{M-2}\right)
...D_{M-1}(\theta ,\theta )$ and $D_N=D_1\left( \theta ,\xi _{N-1}\right)
D_2\left( \theta ,\xi _{N-2}\right) ...D_{N-1}(\theta ,\theta )$, where $%
\theta =\xi _1$, to depict the operate which acts as the following: 
\begin{equation}
D_M\left| \psi _i(\theta )\right\rangle ^{\otimes M}\left| 00\right\rangle
^{\otimes (N-M)}=\left| \psi _i(\xi _M)\right\rangle _{A_1}\left|
00\right\rangle ^{\otimes (N-1)}\text{,}  \eqnum{3.12}
\end{equation}
\begin{equation}
D_N\left| \psi _i(\theta )\right\rangle ^{\otimes N}=\left| \psi _i(\xi
_N)\right\rangle _{A_1}\left| 00\right\rangle ^{\otimes (N-1)}.  \eqnum{3.13}
\end{equation}
where we have defined the vector $\xi _{j+1}$ respectively by 
\begin{equation}
X(\theta ,\xi _j)=\tilde T^{+}(\xi _{j+1})\tilde T(\xi _{j+1}).  \eqnum{3.14}
\end{equation}
As that in Eq.(3.8), we can determine $\xi _{j+1}$ uniquely by above
relation.

For general $M,N$, we still let $\left| P_i\right\rangle =\left|
P_1\right\rangle =\left| 1\right\rangle $, we still denote $\left|
P_0\right\rangle =\left| 0\right\rangle $. For probabilistic identification,
with Eq. (3.12), we can give the orthogonal basis states of Eq. (2.1) as $%
\left\{ \left\{ D_M^{-1}\left| \phi _i\right\rangle _{A_1}\left|
00\right\rangle ^{\otimes M-1}\left| P_0\right\rangle \right\} ,\;\left\{
\left| \phi _j\right\rangle _{A_1}\left| 00\right\rangle ^{\otimes
M-1}\left| P_1\right\rangle \right\} \right. $, $\left. i,j=1,2,3,4\right\} $%
. For probabilistic clone, with Eq. (3.13), we give $\left\{ \left\{
D_M^{-1}\left| \phi _i\right\rangle _{A_1}\left| 00\right\rangle ^{\otimes
N-1}\left| P_0\right\rangle \right\} ,\;\left\{ D_N^{-1}\left| \phi
_j\right\rangle _{A_1}\left| 00\right\rangle ^{\otimes N-1}\left|
P_1\right\rangle \right\} ,\;i,j=1,2,3,4\right\} $.

We introduce two new gates called C-$D_N$-gate and C-$D_N^{+}$-gate, whose
function is that we execute $D_N$ or $D_N^{+}$on the target $N$ states when
the control bit on $\left| 1\right\rangle $, otherwise we make no influence
on the target $N$ states. We can transfer the orthogonal basic states above
into $\left\{ \left| \phi _i\right\rangle _{A_1}\left| 00\right\rangle
^{\otimes M-1}\left| P_j\right\rangle ,\;i=1,2,3,4;\;j=0,1\right\} $ (for
probabilistic identification) and $\left\{ \left| \phi _i\right\rangle
_{A_1}\left| 00\right\rangle ^{\otimes N-1}\left| P_j\right\rangle
,\;i=1,2,3,4;\;j=0,1\right\} $ (for probabilistic clone) with the C-$D_M$
gate, C-$D_N$ gate and $\sigma _x^P$ gate (the inverse with C-$D_M^{+}$
gate, C-$D_N^{+}$ gate and $\sigma _x^P$ gate). On this orthogonal basis
states, we yield 
\begin{equation}
\left\{ 
\begin{array}{c}
\widetilde{V}=V_{A_1}\left| 00\right\rangle ^{\otimes (M-1)\ \otimes
(M-1)}\left\langle 00\right| I_P\text{,\quad for identification} \\ 
\\ 
\widetilde{V}=V_{A_1}\left| 00\right\rangle ^{\otimes (N-1)\ \otimes
(N-1)}\left\langle 00\right| I_P\text{,\quad \quad \quad \quad \quad for
clone}
\end{array}
\right.  \eqnum{3.15}
\end{equation}
where $V_{A_1}$ is a unitary operator on system $A_1$ and its realization
has been resolved by {\sl LEMMA 1}. It is obvious that $\widetilde{V}$ is a
controlled unitary operator.

In the similar way, denote $\hat S_{A_1P}=\sigma _x^1\sigma _x^2\Lambda
_2^{A_1P}(K_1)\sigma _x^2\sigma _x^1\cdot \sigma _x^1\Lambda
_2^{A_1P}(K_2)\sigma _x^1\cdot \sigma _x^2\Lambda _2^{A_1P}(K_3)\sigma
_x^2\cdot \Lambda _2^{A_1P}(K_4)$, where $K_i=\left( 
\begin{array}{cc}
\sqrt{1-m_i} & -\sqrt{m_i} \\ 
\sqrt{m_i} & \sqrt{1-m_i}
\end{array}
\right) $, we can express $S$ as 
\begin{equation}
\left\{ 
\begin{array}{c}
\hat S=\hat S_{A_1P}\left| 00\right\rangle ^{\otimes (M-1)\ \otimes
(M-1)}\left\langle 00\right| ,\quad \text{for identification} \\ 
\\ 
\hat S=\hat S_{A_1P}\left| 00\right\rangle ^{\otimes (N-1)\ \otimes
(N-1)}\left\langle 00\right| ,\quad \quad \quad \quad \quad \text{for clone}
\end{array}
\right.  \eqnum{3.16}
\end{equation}

In the discussion above, we don't set any other limitation of probability
matrix $\Gamma $ except the determining Inequality (2.3) and (2.8). When the
probability changes, the main frame of the networks of probabilistic
identification and clone remains the same but the rotation $\widetilde{V}$
and $\Lambda _2^{A_1P}(K_i)$ have different forms. The stability of the
clone is another virtue of the networks like that in [22]. We can also prove
the capacity of error adaptation of our net. We take the example in FIG.2
and suppose to introduce an error in auxiliary system $B$. The input state
in $B$ comes to 
\begin{equation}
\left| \phi \right\rangle _B=e^{i\tau _1}\sqrt{1-\sum_{i=1}^3\left| \delta
_i^2\right| }\left| \phi _1\right\rangle _B+\sum_{i=1}^3\delta _ie^{i\tau
_{i+1}}\left| \phi _{i+1}\right\rangle _B,  \eqnum{3.17}
\end{equation}
where $\delta _i$ is a variable whose value is close to 0.

Like that in [22], we measure output port $P$. If we get $\left|
0\right\rangle _P$, in general it is regarded as a failure. But if we
measure output following C-$D^{+}$ gate of system $B$ and get $\left| \phi
_{i+1}\right\rangle _B$ which comes from the error $\delta _i$ of $\left|
\phi \right\rangle _B$, it is obvious that the input state won't be
destroyed and can be reset as another input of cloning machine. To the
general situation, when $N$ is relatively large the error happens in the $%
N-M $ ports of $B$ with great probability. If any of the ports comes to this
situation, we can detect it and can avoid incorrect clones and
identification.

\subsection{Realization for $n$-particles System}

In the last part of this section we will discuss the realizing probabilistic
cloning and identifying the states of $n$-particles system. For $n$%
-particles system, there exist $2^n$ linear independent states, which can be
also represented like $\left| \tilde \psi _i\right\rangle $ by a unitary
evolution similar to that in {\sl LEMMA 2}. We can also define the similar
operator $D(\theta ,\xi )$ and $D_M$, $D_N$ that transfer the information of 
$N$ copies into one states of $n$-particles system, where $\theta =\left(
\theta _i^j,\;\mu _i^j,\;i=2,...,2^n;\;j=1,2,...,i-1\right) $. With {\sl %
LEMMA 1 }we can realize $D(\theta ,\xi )$ with universal logic gates. With $%
D(\theta ,\xi )$ operator we can transfer the basis states into the
controlled unitary evolution of $A_1P$ system. We still have

\begin{equation}
U=\widetilde{V}\left( 
\begin{array}{cc}
F & -E \\ 
E & F
\end{array}
\right) \tilde V^{+}\text{.}  \eqnum{3.18}
\end{equation}
where $\widetilde{V}$ is $\left\{ A_i,i=2,3,...,n\right\} $ control $A_1$
unitary operator. We can also express $S$ like Eq.(3.16) with $\sigma
_x^{(i)}$ and $\Lambda _n^{A_1P}(K_j)$ and the controlment of $\left\{
A_i,i=2,3,...,n\right\} $, where $\sigma _x^{(i)}$ represent $\sigma _x$
operator on the $i-$th particle of $A_1$ system. With all the above
operator, we can realize probabilistic cloning and identifying the states of 
$n$-particles system given $M$ initial copies of each state.

\section{CONCLUSION}

We have realize the probabilistic cloning and identifying linear independent
quantum states of multi-particles system, given prior probability, with
universal logic gates using the method of unitary representation, which we
have develope in [22]. In this paper, we don't restrict the states of
multi-particles system are separate or entanglement. So our result is
universal for both of them. We still assume the state of each particle is in 
$2$-level, which means the state can be expressed with $\left|
0\right\rangle $ and $\left| 1\right\rangle $, and the question of how to
realize the probabilistic identifying and cloning the system with $n$-states
of each particle is still interesting.

\quad \quad \quad \quad \quad \quad \quad \quad \quad \quad \quad \quad
\quad {\bf ACKNOWLEDGMENT\ \ }\ \ \ \ \ \ \quad \quad

This work was supported by the National Nature Science Foundation of China.

\newpage\ 

Figure captions:

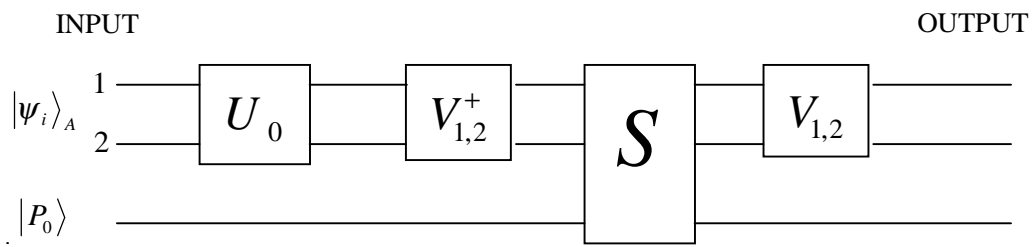
FIG. 1. First we measure the output port $P$. If we get $\left|
0\right\rangle _P$, the identification is successful. Then we measure output
port $1$ and $2$. If it is $\left| \phi _i\right\rangle _{1,2}$, we identify
that the input state is $\left| \psi _i\right\rangle _{1,2}$, whose
probability is $\gamma _i$. If we get $\left| 1\right\rangle _P,$ the
identification fails.

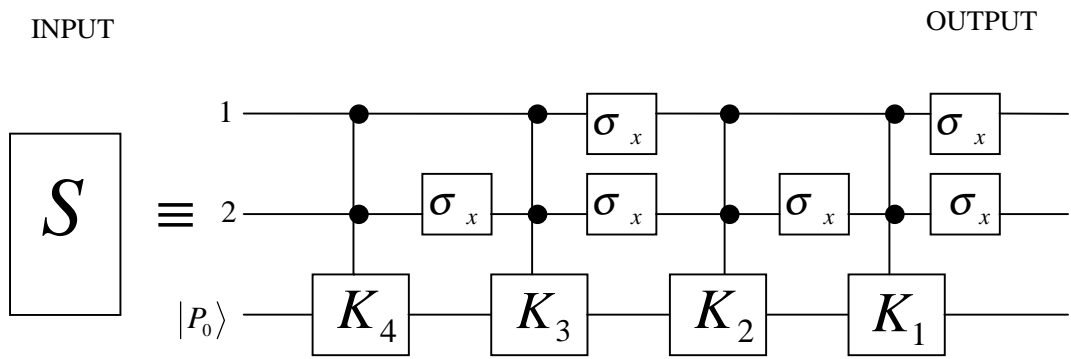
FIG. 2. \ Where $\hat S=\sigma _x^1\sigma _x^2\Lambda _2^{AP}(K_1)\sigma
_x^2\sigma _x^1\cdot \sigma _x^1\Lambda _2^{AP}(K_2)\sigma _x^1\cdot \sigma
_x^2\Lambda _2^{AP}(K_3)\sigma _x^2\cdot \Lambda _2^{AP}(K_3).$

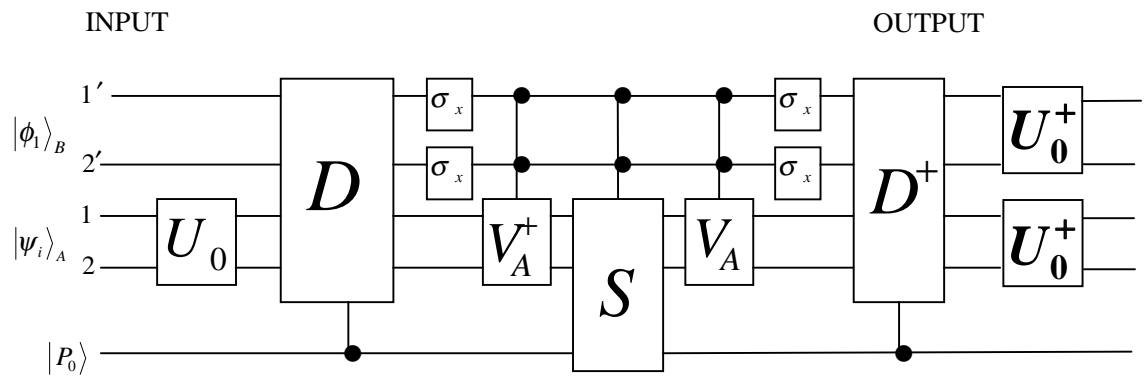
FIG. 3.\quad This quantum networks with logic gates contain a Control-$%
D(\theta ,\theta )$-gate, a Control-$D^{+}(\theta ,\theta )$-gate and a
Control-$S$-gate. We have illustrated $S$ gates in FIG. 2. With the concrete
form of $D(\theta ,\theta )$, $D^{+}(\theta ,\theta )$ and {\sl LEMMA 1 }we
can also express $D(\theta ,\theta )$ and $D^{+}(\theta ,\theta )$ with
quantum logic gates. Here we omit it for it is so complicated and should be
finished by computer. First we measure output port $P$. If we get $\left|
1\right\rangle _P$,the clone is successful. Then two copies of the input
state in input port system $A$ are gained in output port $A$ and $B$. Else
if we get $\left| 0\right\rangle _P$, the clone fails and we discard the
output.

\end{document}